\documentstyle{europhys}
\def\And{{\rm and\ }}

\newif\ifboo \boofalse

\begin{document}
%
%%%   The headers.
%
%%%   These three macros are to have correct headings in your paper.
%%%   You shall omit all the arguments in the two macros `\euro{}{}{}{}'
%%%   `\Date{}' and fill in `\shorttitle{}'. 
%%%   If there is more than one author in the 
%%%   \shorttitle macro, use the macro \etal after first author's name
%%%   to obtain the correct heading.
%
\euro{?}{?}{?}{2004}
\Date{version of 16/04/2004}
\shorttitle{W. WERNSDORFER {\it et al.} Resonant photon absorption ETC.}
\title{Resonant photon absorption in the low spin molecule V$_{15}$}
\author{W. Wernsdorfer\inst{1}, A. M${\rm \ddot u}$ller\inst{2}, 
D. Mailly\inst{3} \And B. Barbara\inst{1}}
\institute{
     \inst{1} Laboratoire Louis N\'eel, associ\'e \`a l'UJF, CNRS, 
     BP 166, 38042 Grenoble Cedex 9, France\\
     \inst{2} Facult${\ddot a}$t f${\ddot u}$r Chemie, Universitat Bielefeld, 
     D-3300501 Bielefeld, Germany\\
     \inst{3} Laboratoire de Photonique et de Nanostructures, 
     CNRS, 91460 Marcoussis, France}
\rec{03 Feb. 2004}{15 April 2004}
\pacs{
\Pacs{75}{45$+$j}{Macroscopic quantum phenomena in magnetic systems}
\Pacs{75}{50 Xx}{Molecular magnets}
\Pacs{75}{60 Ej }{Magnetisation curves, hysteresis, Barkhausen and related effects}
      }
\maketitle
\begin{abstract}
Resonant photon absorption in the GHz range was observed via low
temperature micro-SQUID magnetization measurements 
of the spin ground state $S = 1/2$ of the molecular 
complex V$_{15}$. A simple single-molecule
interpretation is proposed. The line-width essentially results
from intra-molecular hyperfine interaction. The results point out
that observing Rabi oscillations in molecular nanomagnets requires
well isolated low spin systems and high radiation power.
\end{abstract}
%
%%%   Main text
%%%   Sectioning
%%%   In EuroPhys there is only ``one'' level of sectioning `\section{}'.
%

It is widely admitted that single-molecule magnets (SMMs)
have a great potential for quantum computation~\cite{Leuenberger01},
in particular because they are extremely small and almost identical,
allowing to obtain, in a single measurement, statistical averages of
a larger number of qubits. Each spin being carried by a
big molecule of $\sim$ 1 nm$^3$, these magnets are naturally
diluted (with $10^{19}-10^{20}$ molecules/cm$^3$), which is
highly desirable to minimize dipolar interactions and thus
preserve quantum coherence. Despite such natural dilutions,
dipolar interactions remain rather important (0.1 $-$ 0.5 K)
in large spin SMMs such as Mn$_{12}$ or
Fe$_8$ ($S = 10$)~\cite{Christou00} displaying resonant quantum
tunneling of magnetization
~\cite{Thomas96,Friedman96,Sangregorio97,Aubin98,Kent00b,Tasiopoulos04}.
Reductions of dipolar interactions can easily be achieved
by decreasing the value of the spin $S$. At a given concentration,
the dipolar energy scales with the square of $S$.
Thus, decreasing the spin from $S = 10$ to $1/2$
leads to a reduction of dipolar energy by a factor of $(1/20)^2 \sim
10^{-3}$~\cite{ChiorescuV15PRL00}. 
At first glance, low spin systems seem more
suitable for quantum computation than large spin systems.

In this paper we report the first study of the micro-SQUID response
of a low-spin molecular system, V$_{15}$, to electromagnetic radiation.
The advantages of our micro-SQUID technique in respect to
pulsed electron paramagnetic resonance (EPR) techniques
consist in the possibility to perform time-resolved experiments 
(below 1 ns)~\cite{ChiorescuScience03} 
on submicrometer sizes samples (about 1000 spins)~\cite{Jamet01a} 
at low temperature (below 100 mK).
Our first results on  V$_{15}$ open the way 
for time-resolved observations
of quantum superposition of spin-up and spin-down states in SMMs. 
Other results obtained in similar systems but with large spins
concern for example EPR measurements~\cite{Hill98}, 
resonant photon-assisted tunneling 
in a Fe$_8$ SMM~\cite{Sorace03} and non-resonant microwave absorption
in a Mn$_{12}$ SMM~\cite{Amigo03}.

\begin{figure}
\begin{center}
\centerline{\epsfxsize=8 cm \epsfbox{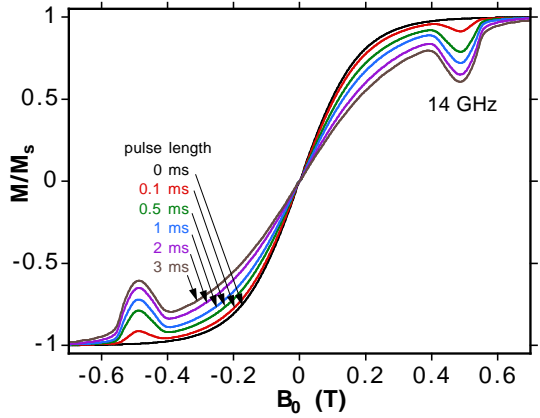}}
\caption{Equilibrium magnetization curves 
measured with or without irradiation. The cryostat temperature 
was 40 mK. The electromagnetic radiation was pulsed with 
a period of 10 ms and a pulse length going from  0 
(no radiation) to 3 ms.}
\label{fig1}
\end{center}
\end{figure}

The molecular complex V$_{15}$ forms a lattice with
space group R$_{3c}$ containing two molecules per
unit cell~\cite{Muller88,Gatteschi93}.
The third order symmetry axis of the unit cell is also the symmetry axis
of the two V$_{15}$ clusters. In each molecule the fifteen V$^{\rm IV}$
ions ($S = 1/2$), are placed on a quasi-spherical layered
structure formed of a triangle sandwiched between
two non-planar hexagons. Each hexagon contains three pairs
of strongly coupled spins and each spin at a corner of the inner
triangle is coupled to two of those pairs (one belonging to the
upper hexagon and one belonging to the lower hexagon)~\cite{Gatteschi93}.
In principle, this molecule should be discussed in terms of the
entanglement of the 15 spins $1/2$, with a Hilbert space
dimension of $2^{15}$. However, considerable simplifications
occur at low temperature where the molecule can be
described by a $S = 1/2$ spin ground state and a $S = 3/2$ excited state.
The energy separation of both spin states of $\approx 3.8 K$
was accurately determined by susceptibility and high-field
magnetization measurements~\cite{Barbara_V15_03}
and by inelastic neutron scattering
experiments~\cite{Chaboussant04}.
Magnetization measurements performed down to 30 mK showed
that the $S=1/2$ spin ground state is split in
zero field by $\approx$ 80 mK~\cite{ChiorescuV15PRL00,Chiorescu03}.
The origin of this splitting for a half-integer spin
is interpreted by the interplay between intra-molecular,
hyperfine and Dzyaloshinskii-Moriya 
interactions~\cite{ChiorescuV15PRL00,Chiorescu03}.
Inter-molecular interactions 
(dipolar or/and residual exchange interactions)
can be evaluated from the low temperature Curie-Weiss law
associated with the spin ground state, giving the Curie
constant $C = 0.686 \mu_{\rm B}$K/T and the paramagnetic
temperature $\theta_{\rm p} \approx 12$ mK,
corresponding to a mean internal field of $\approx$ 12 mT.
This value, one order of magnitude larger than calculated
dipolar interactions is nevertheless much smaller than usual
super-exchange interactions; it is attributed to non-trivial
exchange paths.

\begin{figure}
\begin{center}
\centerline{\epsfxsize=8 cm \epsfbox{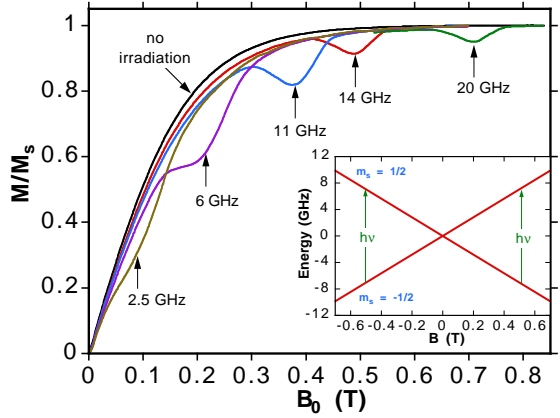}}
\caption{Equilibrium magnetization curves 
similar to Fig. 1 but at several frequencies 
(pulse length 1 ms). Inset: calculated energy levels 
for a spin 1/2 with and $g = 2.02$.}
\label{fig2}
\end{center}
\end{figure}

The measurements were made at a cryostat temperature of 40 mK 
using a 50 $\mu$m sized single
crystal of V$_{15}$. The magnetic probe was a micro-SQUID
array~\cite{WW_ACP_01}
equipped with three coils allowing to apply a field in any
direction and with sweep rates up to 10 T/s.
The electromagnetic radiation was generated by a
frequency synthesizer (Anritsu MG3694A) triggered with a
nanosecond pulse generator.
This setup allows to vary continuously the frequency
from 0.1 Hz to 20 GHz, with pulse lengths form $\sim$1 ns to
1 s~\cite{Thirion03}. The ac radiation field $B_{\rm ac}$ was
directed in a plane perpendicular to the applied static field $B_0$.

Magnetization versus applied field curves $M(B_0)$ were 
measured in the quasi-static regime with a field 
sweep rate slow enough (1 mT/s) to keep the system 
at equilibrium. The phonon-bottleneck regime 
has a characteristic spin-phonon relaxation time 
to the cryostat $\tau_{\rm s}$
of few seconds~\cite{ChiorescuV15PRL00}. 
Ac radiation pulses of 0.1 to 0.3 ms were applied every 10 ms. 
Due to the large value of $\tau_{\rm s}$ the 
relaxation in the intervals between pulses 
is negligible. As a consequence the effects of 
each pulse are additive
leading to an equilibrium magnetization after a time
being larger than $\tau_{\rm s}$.

$M(B_0)$ curves for several pulse lengths, measured at $\nu$ = 14
GHz, are reported in Fig. 1. Two symmetrical dips are clearly
visible at $B_{\nu} = \pm$0.491 T. They result from resonant
absorptions of photons associated with $m_s = 1/2$ to $-1/2$ spin
transitions, as indicated in the inset of Fig. 2. Typical
measurements at other frequencies are presented in Fig. 2. Apart
of the resonant absorptions, a small absorption is also seen at
all applied fields. In order to understand better the origin of
this absorption, Fig. 3 presents the same data as in Fig. 1 but
the magnetization $M$ is converted into a spin temperature $T_s$
using the equation~\cite{Abragam70}:
\begin{equation}
    M(T_s)/M_s = tanh(g\mu_{\rm B}SB_0/k_{\rm B}T_s)
\label{eq_Ts}
\end{equation}
with $S = 1/2$ and  $g = 2.02$ (see below). Fig. 3 shows clearly
the increase of the spin temperature $T_s$ because of resonant
absorptions and a nearly field independent increase of $T_s$.
Before going into a deeper discussion, we show in Fig. 4 the
experimental confirmation of the linear evolution of the resonant
field with frequency $\nu = \gamma B_{\nu}$  predicted by the
inset of Fig. 2. The measured slope $d\nu/dB_{\nu} = \gamma
\approx$ 28.3 GHz/T yields a gyromagnetic ratio $g = 2h \gamma
/\mu_{\rm B} \approx$ 2.02, a value close to the one obtained by
Ajiro et al.~\cite{Ajiro02} in single-crystal EPR measurements
 at 2.4 K  ($g \approx 1.98$).

\begin{figure}
\begin{center}
\centerline{\epsfxsize=8 cm \epsfbox{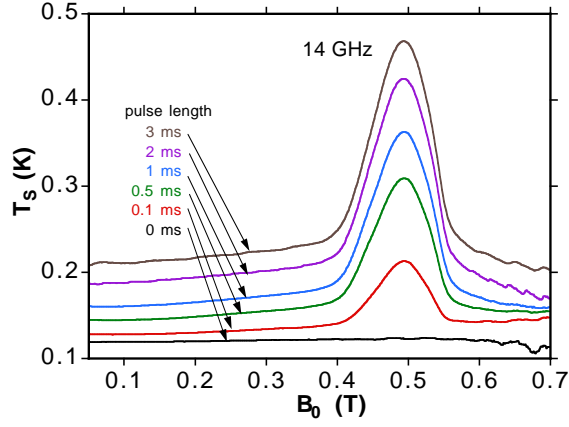}}
\caption{Spin temperature $T_{\rm s}$ 
obtained by inversion of expression (1) where $M(T_{\rm s})$ 
is given by the magnetization curves of Fig. 1. 
The resonant absorption near 0.491 T is superimposed 
to a non-resonant background absorption corresponding 
to a nearly constant spin-temperature.}
\label{fig3}
\end{center}
\end{figure}

\begin{figure}
\begin{center}
\centerline{\epsfxsize=8 cm \epsfbox{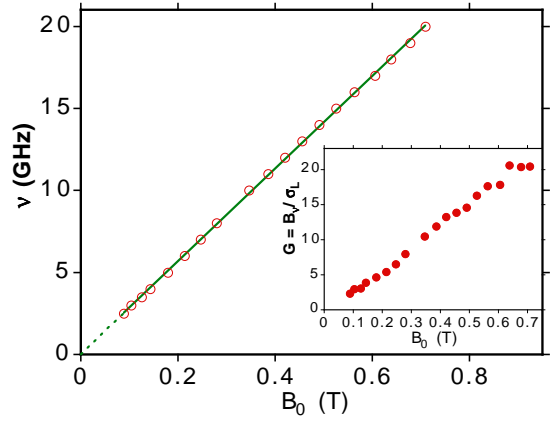}}
\caption{Measured linear variation of the resonance
frequency, corresponding to absorption maxima, with applied magnetic
field. The slope gives $g = 2.02$. Inset: similar variation
obtained for the quality factor $G = B_{\nu}/\sigma_{\rm L}$ representing
a lower bound of the coherence time of molecule spin precession.}
\label{fig4}
\end{center}
\end{figure}

In order to analyze the resonance lines we will consider
the well known Rabi model for a spin $1/2$, which stipulates
that the probability $P$ of finding the system in the
state $m_s= -1/2$ at time $t$, if it was in the state $m_s= 1/2$ at
$t = 0$, is given by~\cite{Rabi37,Abragam70,Grifoni98}:
\begin{equation}
    P = \frac{(\gamma B_{\rm ac})^2}{(\gamma B-\nu)^2+(\gamma B_{\rm ac})^2}
    sin^2(\omega_{\rm Rabi} t)
\label{eq_P_Rabi}
\end{equation}
where $\omega_{\rm Rabi} = 1/2 \sqrt{(\gamma B-\nu)^2+(\gamma B_{\rm
ac})^2}$, $B$ is total field seen by the spin,
and $B_{\rm ac}$ the amplitude of the electromagnetic field.
In the limit of long times,  Eq.~\ref{eq_P_Rabi}
becomes proportional to the time,
$P = \pi (\gamma B_{\rm ac}/2)^2t\delta(\gamma B - \nu)$,
and the transition rate 
$\Gamma = dP/dt = \pi (\gamma B_{\rm ac}/2)^2\delta(\gamma B - \nu)$
turns out to be a constant. Integrating $\Gamma$ over
a distribution of local fields ${\mathcal{F}}(B_{\rm L})$
and considering that $B = B_0 +B_{\rm L}$, where $B_0$
is the applied field, we obtain the field-dependent 
spin-photon transition rate:
\begin{equation}
    \Gamma_{\rm L} = \frac{\pi}{4}\gamma B_{\rm ac}^2 {\mathcal{F}}(B_{\rm
    L} = B_{\nu}- B_0)
\label{eq_Gamma_L}
\end{equation}
This expression shows that the field-dependence of the transition
rate is given by the distribution function of local fields. Below,
we analyze the results in terms of two limiting distributions, a
Gaussian distribution and a Lorentzian one.

The measured magnetization dips corresponding to resonant power
absorption (Figs. 1-3), result from the balance between
the induced transition rate $\Gamma = \Gamma_{\rm L} + \Gamma_0$
and $\tau_{\rm s}$. $\Gamma_0$ is taking into account small heating
effect of the environment.
$\Gamma \tau_{\rm s}$ can be determined using the measured
equilibrium magnetization with and without irradiation,
$M_{\rm RF}(B_0)$ and $M_{\rm 0}(B_0)$ respectively,
and is given by~\cite{Abragam70}:
\begin{equation}
    \Gamma \tau_{\rm s} = \frac{1}{2}
    \biggl(
    \frac{M_{\rm RF}(B_0)}{M_{\rm 0}(B_0)}-1
    \biggr)
\label{eq_Gamma_tau}
\end{equation}

\begin{figure}
\begin{center}
\centerline{\epsfxsize=8 cm \epsfbox{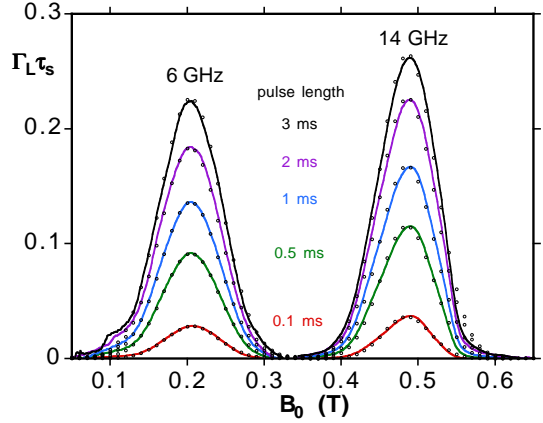}}
\caption{Examples of field dependence 
of $\Gamma_{\rm L}\tau_{\rm s}$  obtained from the 
measured magnetization curves and Eq.~\ref{eq_Gamma_tau}, 
at 6 and 14 GHz, after subtraction of the 
non-resonant absorption background (nearly field independent). 
The dots represent the fit of these curves 
using Eqs.~\ref{eq_Gamma_L} and~\ref{eq_Gauss} (Gaussian distributions)
yielding $\sigma_{\rm L} \approx$ 33 - 38 mT.}
\label{fig5}
\end{center}
\end{figure}

Fig. 5 shows $\Gamma_{\rm L} \tau_{\rm s}$ obtained from the data in
Figs. 1 and 2. Taking the example of the resonance at $\nu$ = 14 GHz, 
$B_0$ = 0.491 T,
we yield $\Gamma_{\rm L} \tau_{\rm s}\approx$ 0.2 (Fig. 5). 
As mentioned above  $\tau_{\rm s} \approx$ 1 s is rather 
large due to the phonon bottleneck effect. This allows to 
obtain sufficiently large values of $\Gamma_{\rm L} \tau_{\rm s}$  
to observe significant radiation absorption even if 
the rate $\Gamma_{\rm L}\approx$ 0.2  $s^{-1}$ is relatively small. 
Using expression (3)we yield $B_{\rm ac} \sim 1 \mu$T. 
Similar values were obtained for all measured frequencies.

The resonant absorption lines of Fig. 5 were well
fit with a Gaussian distribution:
\begin{equation}
    {\mathcal{F}}(B_{\rm L}) = \frac{1}{\sigma_{\rm
    L}\sqrt{2\pi}}e^{-\frac{(B_0 - B_{\nu})^2}{2\sigma_{\rm L}^2}}
\label{eq_Gauss}
\end{equation}
yielding $\sigma_{\rm L} \approx$ 35 mT 
for frequencies between 2 and 20 GHz.

The Gaussian and Lorentzian distributions essentially differ when
$|B_0 - B_{\nu}| > \sigma_{\rm L}$, that is the Gaussian
distribution is more localized (non correlated fields) than the
Lorentzian one (correlated fields). Our result suggests therefore
that the line-broadening is due to fluctuations of non-correlated
intra-molecular nuclear spins and not to correlated
inter-molecular interactions.

This is confirmed by several features: 

(i) $\sigma_{\rm L}
\approx$ 35 mT (Fig. 5) is much
larger than the mean dipolar field seen by the molecules ($<$ 1 mT);

(ii) $\sigma_{\rm L}$ is also larger than intermolecular
interactions (12 mT);

(iii) $\sigma_{\rm L}$ is close to the hyperfine coupling
in V$_{15}$ ($\approx$ 40 mT)~\cite{ChiorescuV15PRL00,Barbara_V15_03,Chiorescu03}. This
value is derived from the hyperfine field of 11.2 T/$\mu_{\rm B}$
obtained in different systems of V$^{\rm IV}$ ~\cite{Carter77} and
giving a field of about 20 mT for a single V$^{\rm IV}$ spin;
% and
% of $20\sqrt{15}/2$ mT = 38 mT for the whole 
% molecule spin 1/2, considered
% as a central spin;

(iv) finally, $\sigma_{\rm L}$ is
very close to the level width (40 mK) obtained from
magnetic relaxation experiments near the $m_s= \pm1/2$
avoided level crossing.
Its origin was mainly attributed to hyperfine 
and weak Dzyaloshinskii-Moriya interactions
~\cite{ChiorescuV15PRL00,Barbara_V15_03,Chiorescu03}.

This hyperfine broadening of resonant absorption is
consistent with theoretical investigations according which
the main dephasing mechanism of magnetic molecules
is connected with nuclear 
spins~\cite{GargPRL95,Prokofev96,Prokofev98,Garanin00,Dobrovitski03,Stamp03}.

The inset of Fig. 4 presents the field dependence of the quality
factor $G = B_{\nu}/\sigma_{\rm L}$ that gives a lower bound of the
precessional coherence of V$_{15}$ molecule spins. A linear
increase of $G$ is observed reaching $G$ = 20 at 0.7 T that is
rather far from values of $10^4$ needed for quantum computation.

Finally, we discuss the possibility of observing Rabi oscillations
with the present set-up. 
Due to inhomogeneous broadening only a low bound of 
the coherence time can be estimated 
from the resonance lines: $\tau_{\rm c}^{-1} = \gamma\sigma_{\rm L}$. 
The corresponding number of coherent flips of the 
spin system is given by $N = \tau_{\rm c}/\tau_{\rm Rabi} = B_{\rm
ac}/(2\pi\sigma_{\rm L})$  (Rabi coherence). Using the values of
$B_{\rm ac}$ and $\sigma_{\rm L}$ obtained from the fit of the
resonance lines, we get $\tau_{\rm c} \approx$ 2 ns 
(comparable to recent EPR measurements~\cite{Hill03}) and $N \approx
10^{-6}$, showing that there is no hope to see Rabi oscillations
in the present conditions. In order to get $N >> 1$, it will be
necessary to increase the radiation field $B_{\rm ac}$ by orders of
magnitude. Note that $\Gamma_{\rm L,Max} \approx \gamma B_{\rm
ac}^2/4\sigma_{\rm L}$ increases with electromagnetic power and
decreases for broader distribution of local fields.
An important consequence of intra-molecular 
inhomogeneous broadening is the possibility 
to significantly longer coherence times that can be evidenced
using spin-echo techniques. 
This should lead to a transverse relaxation time 
$\tau_2 >> \tau_{\rm c}$ which will be necessary for quantum computation.

In conclusion, we presented a new technique for studying
radiation absorption in the molecular
system V$_{15}$ constituting a first step towards the observation
of Rabi oscillations in molecular magnets.  The main results are
the observation of relatively narrow resonant absorption lines
that are dominated by hyperfine interaction. In order to observe
Rabi oscillations in a magnetic system, an important requirement is
a large ac field amplitude.

This work was supported by CNRS, Rhone-Alpe,
and the European Union TMR network MOLNANOMAG,
HPRN-CT-1999-0012.

% Create the reference section using BibTeX:
%\bibliography{basename of .bib file}
%\bibliographystyle{wernsdor}
%\bibliography{wernsdor}

\end{document}